\documentclass[twocolumn,english,aps,prd,longbibliography,groupedaddress]{revtex4-1}
\usepackage[T1]{fontenc}
\usepackage[utf8]{inputenc} 
\usepackage{color}
\usepackage{bm}
\usepackage[linesnumbered,ruled,vlined]{algorithm2e}
\usepackage{xcolor}
\usepackage{amsmath}
\usepackage{mathtools}
\usepackage{amssymb}
\usepackage{graphicx}
\usepackage{natbib}
\usepackage{braket}
\usepackage{psfrag}
\usepackage{tikz}
\usepackage[normalem]{ulem}
\usepackage{qcircuit}
\usepackage{siunitx}

\DeclarePairedDelimiter\floor{\lfloor}{\rfloor}

\usepackage{url}

\usepackage[colorlinks=true,urlcolor=blue,citecolor=blue,linkcolor=blue]{hyperref}
\pdfmapfile{+sansmathaccent.map}

\newcommand\numberthis{\addtocounter{equation}{1}\tag{\theequation}}

\begin{document}
 

\title{Optimizing Quantum Search Using a Generalized Version of Grover's Algorithm}
 
\author{Austin Gilliam}
\affiliation{
JPMorgan Chase}

\author{Marco Pistoia}
\affiliation{
JPMorgan Chase}

\author{Constantin Gonciulea}
\affiliation{
JPMorgan Chase}
 
 
\date{\today}
 
\begin{abstract}
Grover's Search algorithm was a breakthrough at the time it was introduced, and its underlying procedure of amplitude amplification has been a building block of many other algorithms and patterns for extracting information encoded in quantum states. 
In this paper, we introduce an optimization of the inversion-by-the-mean step of the algorithm.  This optimization serves two purposes: from a practical perspective, it can lead to a performance improvement; from a theoretical one, it leads to a novel interpretation of the actual nature of this step. This step is a reflection, which is realized by (a) cancelling the superposition of a general state to revert to the original all-zeros state, (b) flipping the sign of the amplitude of the all-zeros state, and finally (c) reverting back to the superposition state. Rather than canceling the superposition, our approach allows for going forward to another state that makes the reflection easier.  We validate our approach on set and array search, and confirm our results experimentally on real quantum hardware. 
\end{abstract}
 
\maketitle
 

\section{\label{sec:introduction}Introduction}
The capabilities of quantum computers are evolving at a very fast pace.  With half a century of research efforts on theoretical quantum computing, increasingly more researchers are now working on designing and implementing new quantum algorithms that can take advantage of the fast-evolving underlying quantum hardware.  While some of such algorithms are meant to address a particular problem in a specific domain, a large body of research has been devoted to the creation of algorithms of general applicability, such as Shor's algorithm~\cite{shor1999polynomial}, Grover's Search~\cite{Grover1996}, and Variational Quantum Eigensolver~\cite{VQE}.

Grover's Search algorithm is of particular interest to researchers due to its vast area of applicability, which spans across multiple domains. 
With high probability, Grover's Search finds an output of interest in an unstructured search space with quadratic speedup compared to classical solutions.
The algorithm, introduced by Lov Grover in 1996, has been expanded on several times since its first formulation~\cite{Boyer1998,Mosca1998,Brassard2000}.

In this paper, we propose further optimizations of Grover's Search algorithm, as follows:
\begin{enumerate}
    \item A generalization of the inversion-by-the-mean step,
    \item A modified version of the original algorithm formulation, which we describe as \emph{set search}, and
    \item A modification of \emph{array search}, which we had introduced in previous work~\cite{Gonciulea2019}.
\end{enumerate}
To the best of our knowledge, these contributions are novel.  Additionally, we demonstrate experimentally, on real quantum hardware, that these contributions can lead to more optimal realizations of quantum search.

The modified Grover iterate presented in this paper applies to the search, counting and optimization features of the quantum-dictionary structure~\cite{Gonciulea2019}, as well as to other algorithms that use amplitude amplification.

The remainder of this paper is organized as follows: Section \ref{sec:grover} offers an overview of the standard formulation of Grover's Search algorithm.
Section \ref{sec:grover-gen} shows how our generalization and modification of Grover's Search algorithm can be applied to set search and array search.  
Section \ref{sec:experimental-results} demonstrates an experimental validation of our approach on real quantum hardware.  
Related work around modifications and extensions of Grover's Search algorithm is covered in Section \ref{sec:related}.
Finally, Section \ref{sec:conclusion} concludes the paper and discusses potential future directions for this work.

\section{\label{sec:grover}Standard Grover's Search}
Grover's Search algorithm~\cite{Grover1996} was created in the context of unstructured search, where we assume a single state of interest. The algorithm is summarized below:
\begin{enumerate}
    \item \label{grover-step-1} Initialize a quantum system of $n$ qubits to a state of equal superposition.
    \item \label{grover-step-2} Repeat the following steps $O(\sqrt{2^n})$ times:
    \begin{enumerate}
        \item Apply an oracle $O$, which recognizes the state of interest and multiplies its amplitude by $-1$.
        \item \label{grover-step-2b} Apply an operator that performs an inversion by the mean on all amplitudes. This is typically done by removing the superposition, multiplying the amplitude of the $\ket{0}_n$ state by -1, and then restoring the superposition.
    \end{enumerate} 
\end{enumerate}
Step~\ref{grover-step-2} describes the central concept of quantum search. 
When applied a precise number of times, it incrementally amplifies the magnitude of the amplitudes of the states of interest, thus increasing their probabilities of being measured.

Geometrically, the multiplication of the amplitude of the $\ket{0}_n$ state by $-1$ is a reflection, which we will denote by $M_0$ for \emph{mirror}---a convention used in Geometric Algebra and other literature as well~\cite{OReillyBook}.  For a given state vector $s = \sum_{j = 0}^{2^n-1} \alpha_j \ket{j}$, we define:

\begin{eqnarray}
M_0(s) := \sum_{j \neq 0} \alpha_j\ket{j}_n - \alpha_0\ket{0}_n
\label{eq:diffusion}
\end{eqnarray}

In his original paper~\cite{Grover1996}, Grover referred to Step~\ref{grover-step-2b} as \emph{diffusion} and the mirror operation as a \emph{rotation}.
Note that some authors indicate by \emph{diffusion} only the reflection in $\ket{0}_n$, which is why we choose to use the term \emph{mirror}.

The Hadamard operation is used to create and revert from superposition.

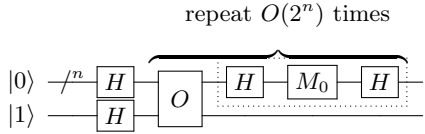
\begin{figure}[ht]
\begin{equation*}
    \Qcircuit @C=1.0em @R=0.0em @!R {
    	& & & & \mbox{{\color{white}aaaaaaa}repeat $O(2^n)$ times} \\
    	\\
	 	\lstick{\ket{0}} & \qw{/^n}  & \gate{H} & \multigate{1}{O}   & \gate{H} & \gate{M_0} & \gate{H} & \qw \\
	 	\lstick{\ket{1}} & \qw      & \gate{H} & \ghost{O}          & \qw & \qw      & \qw      & \qw
	 	\gategroup{3}{5}{3}{7}{.7em}{.}
	 	\gategroup{3}{4}{3}{7}{.7em}{^\}}
	 }
\end{equation*}
\caption{\label{fig:simple-grover}A circuit representing the simplest case of Grover's Search algorithm.}
\end{figure} 

The algorithm was later generalized, allowing for any unitary state-preparation operator $A$ (i.e., not necessarily an equal superposition created with the Hadamard operator $H$) and multiple marked states (sometimes called the \emph{good states}), and is now commonly referred to as \emph{amplitude amplification}~\cite{Brassard2000}. 
Step~\ref{grover-step-2}, known as the \emph{Grover iterate}, takes the form $G=-AS_0A^{\dagger}O$, where $S_0$ is the same operator we denote by $M_0$.
Note that the negative sign can be ignored in the implementation, leading to a small adjustment in the interpretation of measurements.
Note also that in some literature, the combination $A^{\dagger}OA$ is referred to as the \emph{oracle}, but we prefer to keep the definitions of $A$ and $O$ separate.

\section{\label{sec:grover-gen}Variation of Grover's Search}

In this paper, we present a more general form the Grover iterate $G=B^{\dagger}M_{B}BO$, where we use operator $B$ and mirroring operator $M_{B}$, which depends on $B$.
As we will show, if the implementations of $B$ and $M_B$ are efficient, this generalization can lead to more optimal realizations of quantum search. 
We retrieve the known form of the Grover iterate by taking $B=A^{\dagger}$ and $M_{B}=M_0$. 

\begin{figure}[ht]
\begin{equation*}
    \Qcircuit @C=1.0em @R=0.7em {
	 	\lstick{\ket{0}^0{\color{white}aa}} & \gate{X} & \ctrl{4} & \gate{X} & \qw \\
	    \lstick{\ket{0}^1{\color{white}aa}} & \gate{X} & \ctrl{3} & \gate{X} & \qw \\
	 	\lstick{\vdots{\color{white}aa}} & \vdots & & \vdots \\
	 	\\
	 	\lstick{\ket{0}^{n-2}} & \gate{X} & \ctrl{1} & \gate{X} & \qw \\
	 	\lstick{\ket{0}^{n-1}} & \gate{X} & \gate{Z} & \gate{X} & \qw
	 }
\end{equation*}
\caption{\label{fig:mirror}A circuit implementation of the mirror operator $M_0$.}
\end{figure}
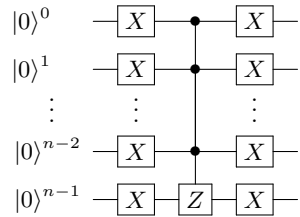

We will explore a particular case of this generalization, as an optimized alternative to the standard implementation of Grover's Search, which assumes that before applying the mirror operator we must be in the state $\ket{0}_n$.
The mirror operation is implemented as $M_0=X^{\otimes n}M_1X^{\otimes n}$, where $M_1$ is the operator that flips the sign of the all-ones state $\ket{{2^n-1}}_n$ i.e., for a given state vector $s = \sum_{j = 0}^{2^n-1} \alpha_j \ket{j}$, we have: 

\begin{eqnarray}
M_1(s) = \sum_{j \neq 2^n-1} \alpha_j\ket{j}_n - \alpha_{2^n-1}\ket{{2^n-1}}_n
\end{eqnarray}
It is easy to verify that the circuit in Figure~\ref{fig:mirror} is an implementation of $M_0$.

As an algebraic intuition, if we can combine the action of the $X$ gates with the previous operator into a more efficient operator, the result can be more efficient overall.

More formally, if we have an operator $B$ such that $BA=X^{\otimes n}$, then $A^{\dagger}=X^{\otimes n}B$ and $A=B^{\dagger}X^{\otimes n}$, and the Grover iterate becomes:
\begin{align*}
G &= AM_0A^{\dagger}O \\
  &= B^{\dagger}X^{\otimes n}M_1X^{\otimes n}BO \\
  &= B^{\dagger}X^{\otimes n}X^{\otimes n}M_1X^{\otimes n}X^{\otimes n}BO \\
  &= B^{\dagger}M_1BO \numberthis
\end{align*}
In this case, $M_B=M_1$, which uses fewer gates than $M_0$, as we avoid the $X$ gates on either side of the controls.

\subsection{\label{sec:grover-set}Set Search}
The context in which Grover's Search algorithm was originally introduced can be described as a \emph{set search}, where we are looking for states of interest in an unstructured collection of data. In this context, the modified version of the algorithm uses $A=R_X(\frac{\pi}{2})$, $B=-iR_X(\frac{\pi}{2})$, and $M_B=M_1$. In many cases, we can ignore the additional rotation added by the $-i$ factor, and use $B=R_X(\frac{\pi}{2})$.

The modification results in a reduction in the total number of gates used in quantum computations that utilize a Grover iterate---which can potentially lead to better overall performance, as confirmed experimentally in Section~\ref{sec:grover-set-results}.

\subsection{\label{sec:grover-list}Array Search}
Many presentations of Grover's Search algorithm focus on the set search version, where values are put in superposition in a single register before the amplitudes of the desired outcomes are amplified.
However, the real power of quantum search is revealed when multiple entangled registers are used.

To show that, we will consider the case of an array search, where values are indexed by a separate register.
In general, we do not know how many values are present, or how many times a single value is repeated.
In such situations, one can use \emph{quantum counting} first, revealing the number of times one needs to apply the Grover iterate to find one of the desired values and its index.
In~\cite{Gonciulea2019}, we show how to implement quantum search and counting on a quantum dictionary, a pattern for representing key/value pairs as entangled quantum registers. Alternatively, we can use adaptive versions of Grover's algorithm that randomize the number of times the Grover iterate is applied.

In the method described in~\cite{Gonciulea2019, Gilliam2019}, the values in the array are encoded using an operator of the form $A=PH$, which first puts all indices and possible values in superposition, and then entangles each array index with a corresponding value.

For a given value of interest (in the context of counting or searching), we build a Grover iterate of the form $G=PHM_0HP^{\dagger}O$.

The modified version uses operators $A=PR_X(\frac{\pi}{2})$ and $B=R_X(\frac{\pi}{2})P^{\dagger}$ (ignoring the $-i$ factor in the cases it can be done), and mirror operator $M_B=M_1$, such that the Grover iterate becomes $G = PR_X(-\frac{\pi}{2})M_1R_X(\frac{\pi}{2})P^{\dagger}O$.

\begin{figure}[ht]
\includegraphics[width=8cm]{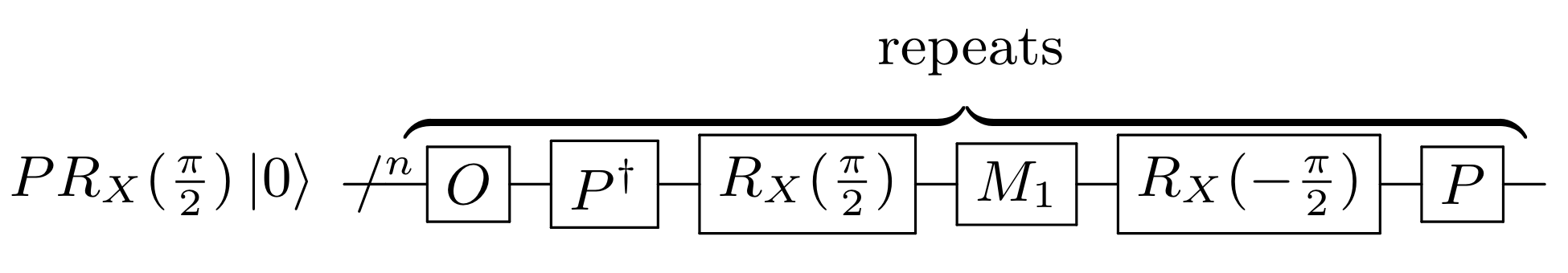}
\caption{\label{fig:grover-iterate-array-search}A circuit implementation of the Grover iterate $G$ used in array search.}
\end{figure} 


A description of $P$ is included in Appendix~\ref{sec:quantum_dictionary}, and an example circuit for $A$ is shown in Section~\ref{sec:grover-list-results}.

\subsection{\label{sec:other-applications}Other Applications}
The ideas presented in Section~\ref{sec:grover-list} are equally applicable to the adaptive versions of Grover's Search algorithm and Amplitude Estimation.

As a general rule, the usages of $R_X(\pi/2)$ within this paper can be replaced by $R_Y(\pi/2)$---which has a counterpart in the world of probabilistic bits.  Thus, the method can also be applied to a probabilistic version of Grover's Search algorithm.

\section{\label{sec:experimental-results}Impact of Gate-Count Reduction: Experimental Results}

In this section, we take a closer look at concrete applications of this generalization, including analyses of performance on real quantum computers.

\subsection{\label{sec:grover-set-results}Set Search}
\begin{figure}[ht]
\includegraphics[width=7cm]{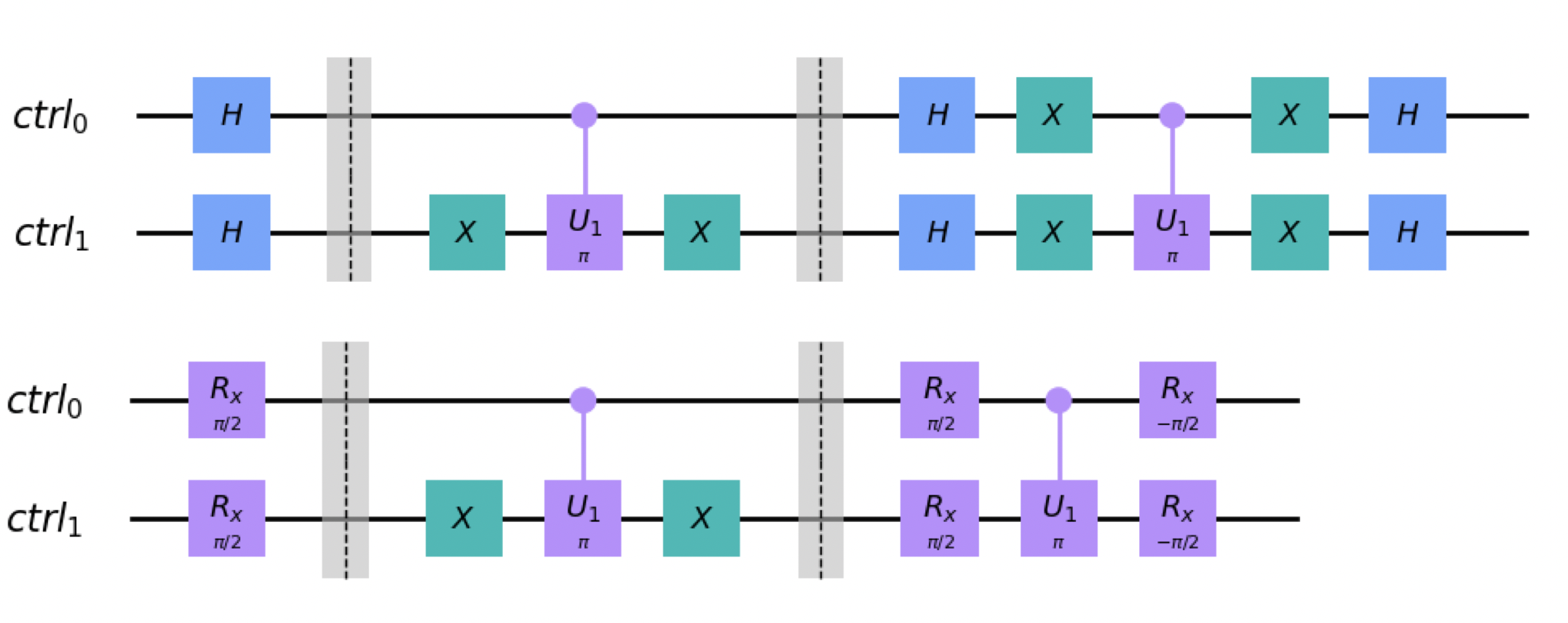}
\caption{\label{fig:grover-circuit-set} Circuits for the \emph{standard} (top) and \emph{modified} (bottom) version of Grover' Search, searching for $2$ with $n=2$ qubits. Circuits are generated using Qiskit~\cite{Qiskit}.}
\end{figure}

Let us compare the standard version of Grover's Search ($A=H$) with the modified version for $n=2$ qubits, using $2$ (\textbf{10} in binary representation) as our state of interest (Figure~\ref{fig:grover-circuit-set}). 

 The result is a reduction in the number of $X$ gates, as seen in the table below:

\begin{center}
\begin{tabular}{ c | c c c c }
 Implementation  & $H$ & $R_X$ & $X$ & $CU_1$ \\ 
 \hline
 Standard & 6 & 0  & 6 & 2 \\ 
 Modified & 0 & 6  & 2 & 2
\end{tabular}
\end{center}
The reduction in the gate count continues for $n=3$ qubits, as shown below. 
In general, we save $2n$ $X$ gates per mirror operator, and we apply the mirror operator $\floor{\pi\frac{\sqrt{2^n}}{4}}$ times~\cite{Boyer1998}.

\begin{center}
\begin{tabular}{ c | c c c c }
 Implementation  & $H$ & $R_X$ & $X$ & $CCU_1$ \\ 
 \hline
 Standard & 15 & 0 & 16 & 4 \\ 
 Modified & 0 & 15 & 4  & 4 
\end{tabular}
\end{center}

\subsection{\label{sec:grover-list-results}Array Search}
\begin{figure}[ht]
\includegraphics[width=4cm]{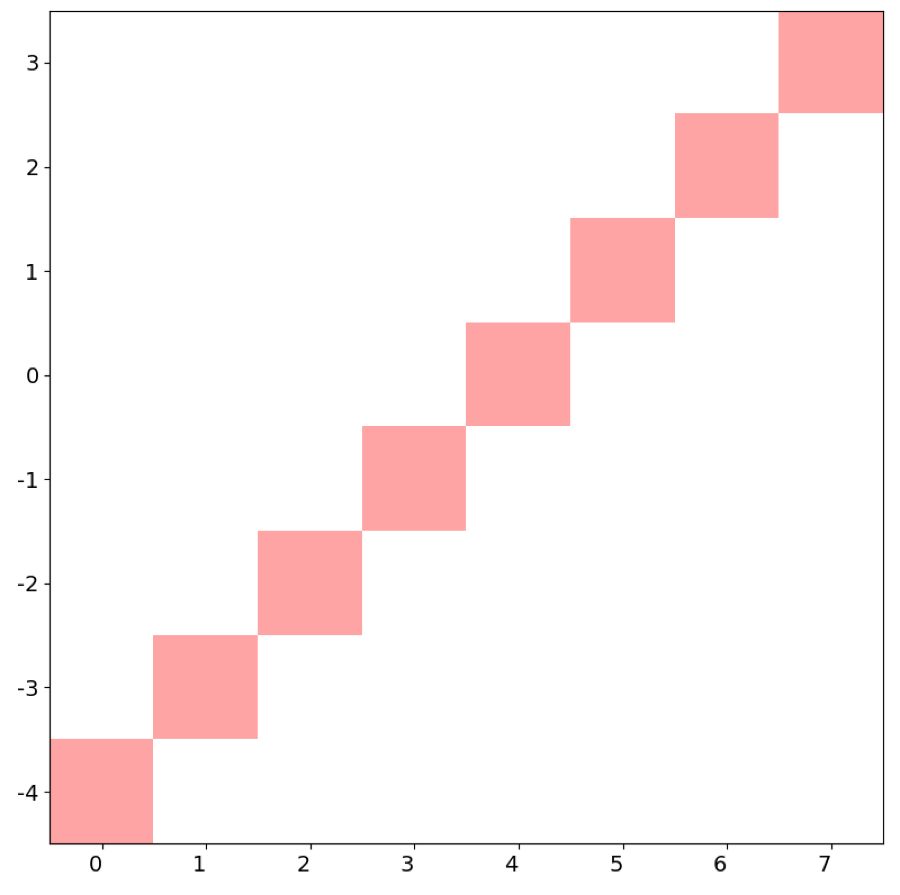}
\caption{\label{fig:ae-example-pixel-graph}A visual representation of the quantum state encoding the array described in Section~\ref{sec:grover-list}. The indices are represented on the horizontal axis, and the values on the vertical.}
\end{figure}

Consider the array $[-4, -3, -2, -1, 0, 1, 2, 3]$, whose values are chosen to match the polynomial formula $f(j)=j-4$ for integer indices $0 \le j \le 7$. 
An efficient way to encode polynomial values in a quantum register is described in~\cite{Gonciulea2019, Gilliam2019}.
A visual representation of the array is given in Figure~\ref{fig:ae-example-pixel-graph}, where the index register is shown on the horizontal, and the value register on the vertical.
Each pixel represents the amplitude of the index/value pair, where the intensity represents the magnitude of the amplitude, and the color is determined by its phase.
A more detailed description can be found in Appendix~\ref{sec:pixel_based_visualization}.

\begin{figure}[ht]
\includegraphics[width=8cm]{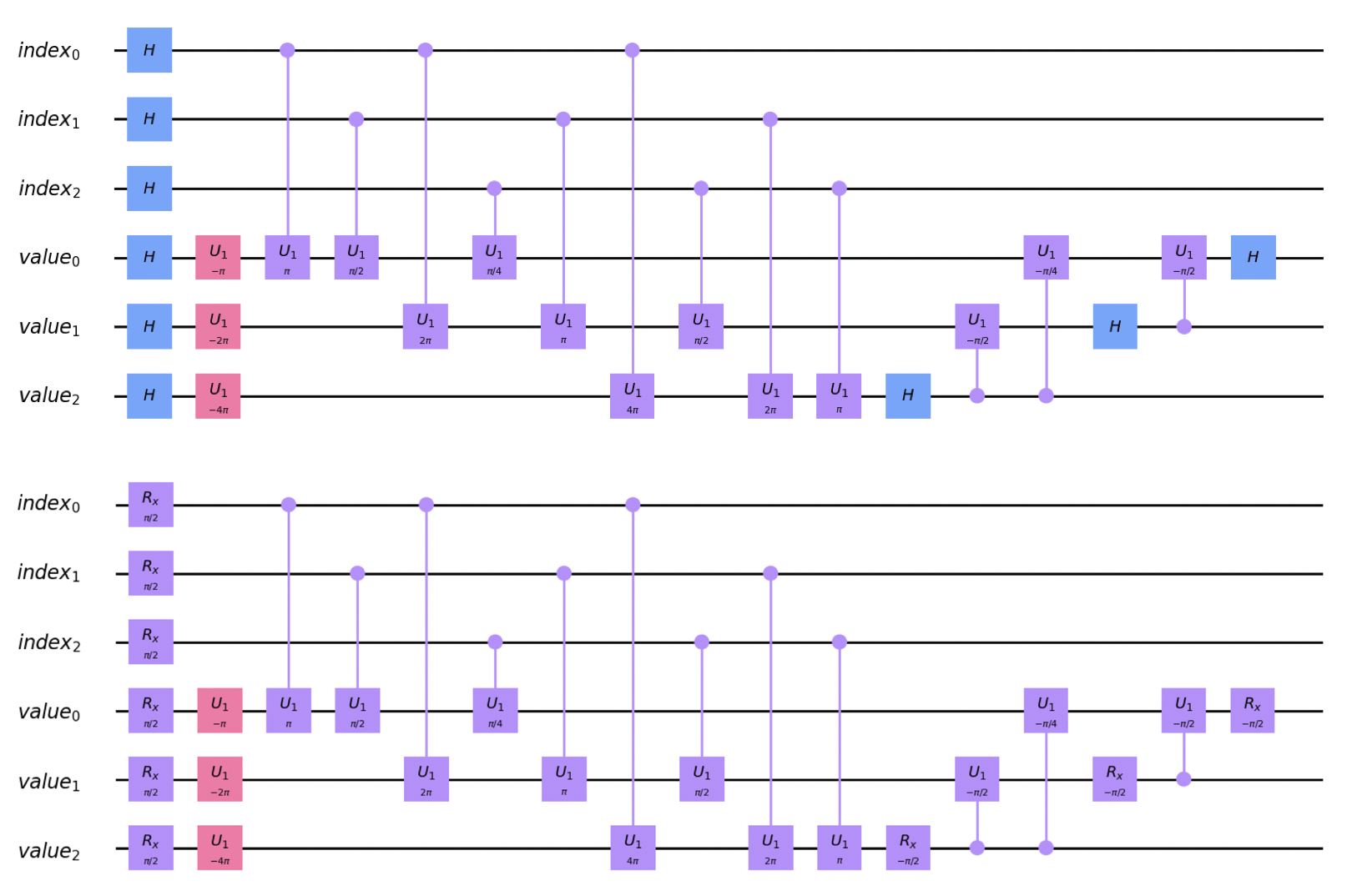}
\caption{\label{fig:aa-encoding-circuits}The circuits representing the operator $P$ for the \emph{standard} (top) and \emph{modified} (bottom) versions of array search. Circuits are generated using Qiskit~\cite{Qiskit}.}
\end{figure}

The circuits for encoding such a state with the standard and modified versions are shown in Figure~\ref{fig:aa-encoding-circuits}.

\begin{figure}[ht]
\includegraphics[width=8cm]{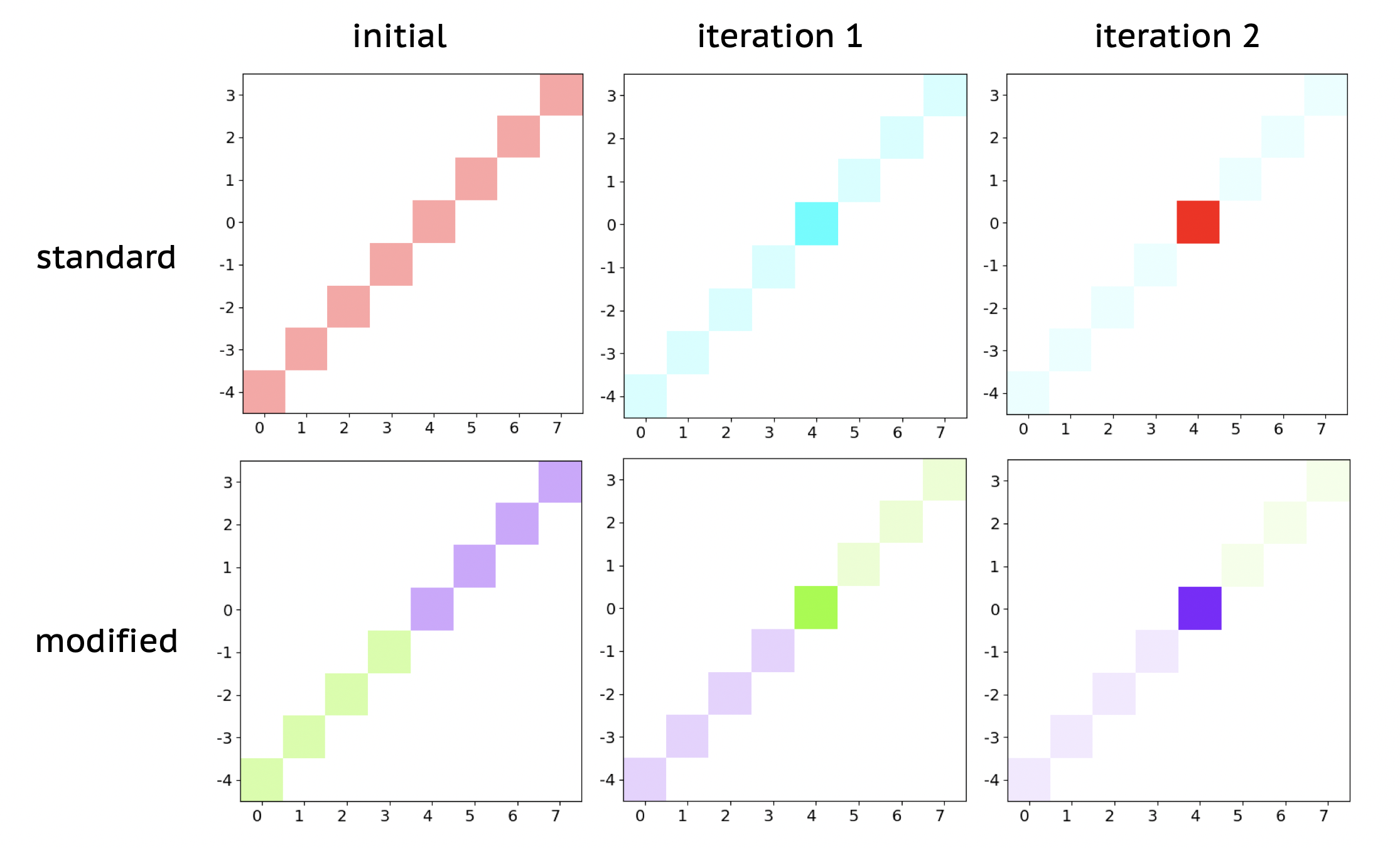}
\caption{\label{fig:aa-results}A visual representation of each iteration of array search, using both the standard and modified versions.}
\end{figure} 

The modified version reduces the number of $X$ gates, as seen in the table below. Note that $nCU_1$ denotes an n-controlled $U_1$ gate, where $n+1$ is the number of index and value qubits.

\begin{center}
\begin{tabular}{ c | c c c c c c}
 Implementation  & $H$ & $R_X$ & $X$ &$U_1$ & $CU_1$ & $nCU_1$ \\ 
 \hline
 Standard & 45 & 0  & 36 & 15 & 60 & 4 \\ 
 Modified & 0  & 45 & 12 & 15 & 60 & 4
\end{tabular}
\end{center}
The results for both the standard and modified versions are shown in Figure~\ref{fig:aa-results}, which visualizes each iteration of the search. Note that using $R_X(\frac{\pi}{2})$ leads to different amplitude phases, represented in the color of each pixel.

\subsection{\label{sec:grover-set-hardware-results}Real Hardware}
As seen in Sections \ref{sec:grover-set-results} and \ref{sec:grover-list-results}, the modified version of Grover's Search decreases the gate count of the circuit. 
In this section, we present how this affects results on current quantum hardware. 
Note that the results of any compiled circuit is dependent on the performance of circuit transpilers, which may theoretically make the proposed optimization already.
However, it is beneficial to make these optimizations by design, as we show here.
Note also that results may differ day by day (due to hardware tuning) and by device, so while we will not always see the same results, a lower gate count offers the potential for improvement.
For details on the quantum hardware used in the following examples, see Appendix~\ref{sec:hardware_config}.

\begin{figure}[ht]
\includegraphics[width=7cm]{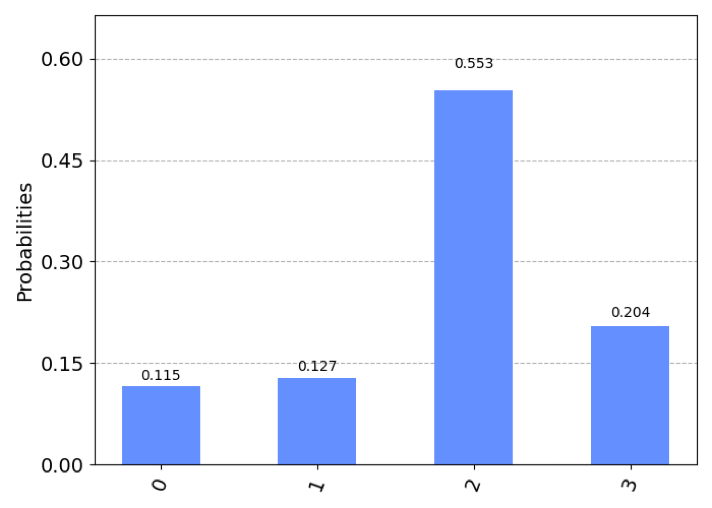}
\caption{\label{fig:grover-2-qubit-set-h}The aggregated result of running the same circuit three times on IBM's \texttt{ibmq\_burlington} backend with 8192 shots, using the \emph{standard} version with $n=2$ qubits, searching for the value $2$.}
\end{figure}

\begin{figure}[ht]
\includegraphics[width=7cm]{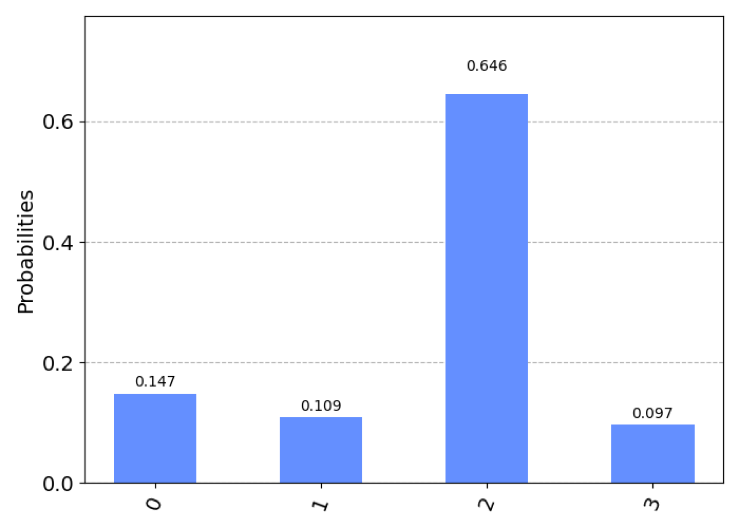}
\caption{\label{fig:grover-2-qubit-set-rx}The aggregated result of running the same circuit three times on IBM's \texttt{ibmq\_burlington} backend with 8192 shots, using the \emph{modified} version with $n=2$ qubits, searching for the value $2$.}
\end{figure}

Returning to the two-qubit examples given in Section~\ref{sec:grover-set-results}, the two versions perform similarly on a high-fidelity quantum computer. However, as the coherence times decrease, the modified version has a notable advantage, as seen in Figures~\ref{fig:grover-2-qubit-set-h} and~\ref{fig:grover-2-qubit-set-rx}.

We expect the desired value (in this case, $2$) to have the highest probability, and stand out among the other outcomes.
This is clearly more pronounced with the modified version (see Figure~\ref{fig:grover-2-qubit-set-rx}), compared to the standard (see Figure~\ref{fig:grover-2-qubit-set-h}.
Running the experiment multiple times, the modified version resulted in the desired value having a probability that was 15\% higher than the one in the standard version.

\begin{figure}[ht]
\includegraphics[width=7cm]{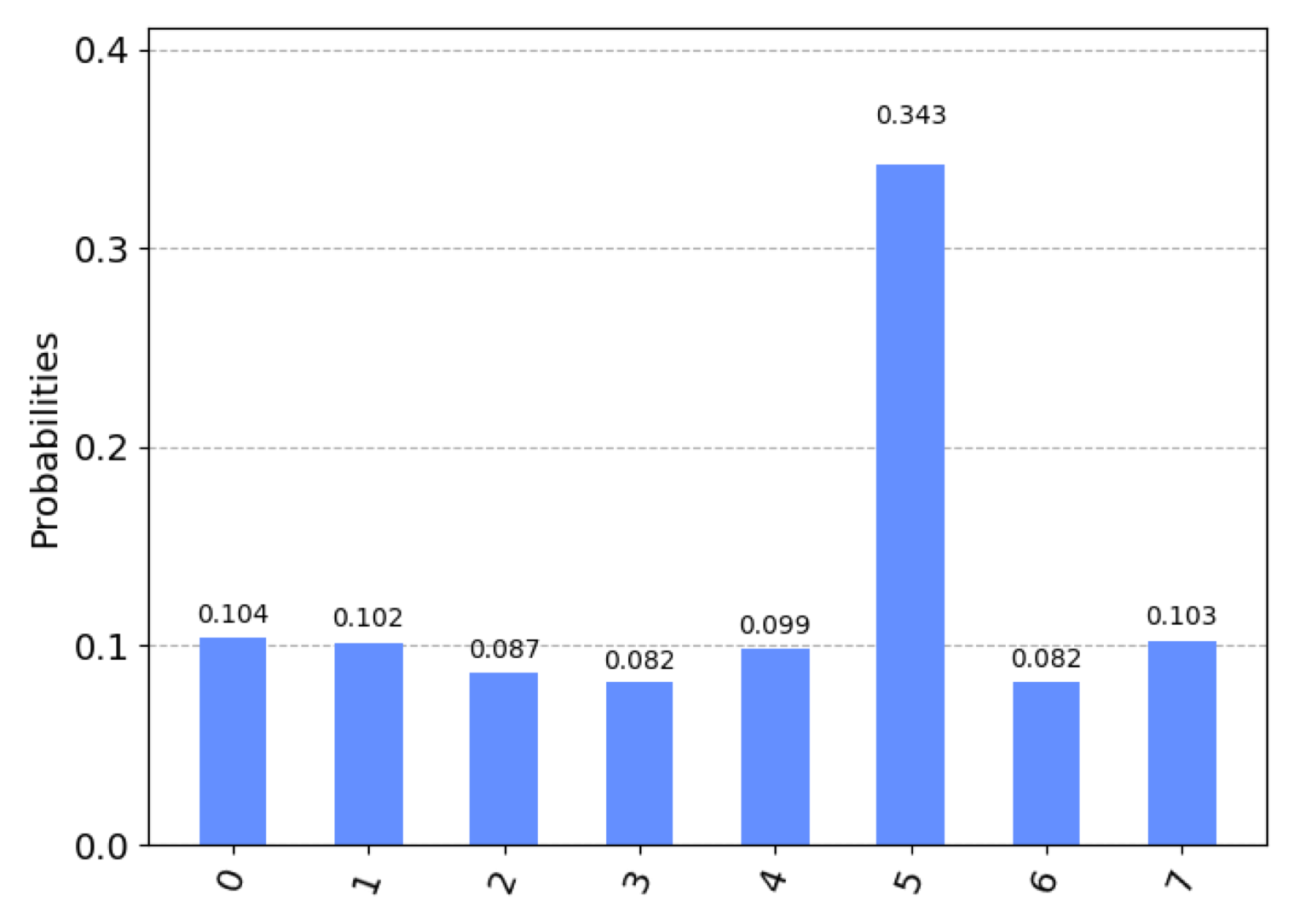}
\caption{\label{fig:grover-3-qubit-set-h}The result of running a Grover circuit on IBM's \texttt{ibmq\_ourense} backend with 8192 shots, using the \emph{standard} version with $n=3$ qubits, searching for $5$.}
\end{figure} 

\begin{figure}[ht]
\includegraphics[width=7cm]{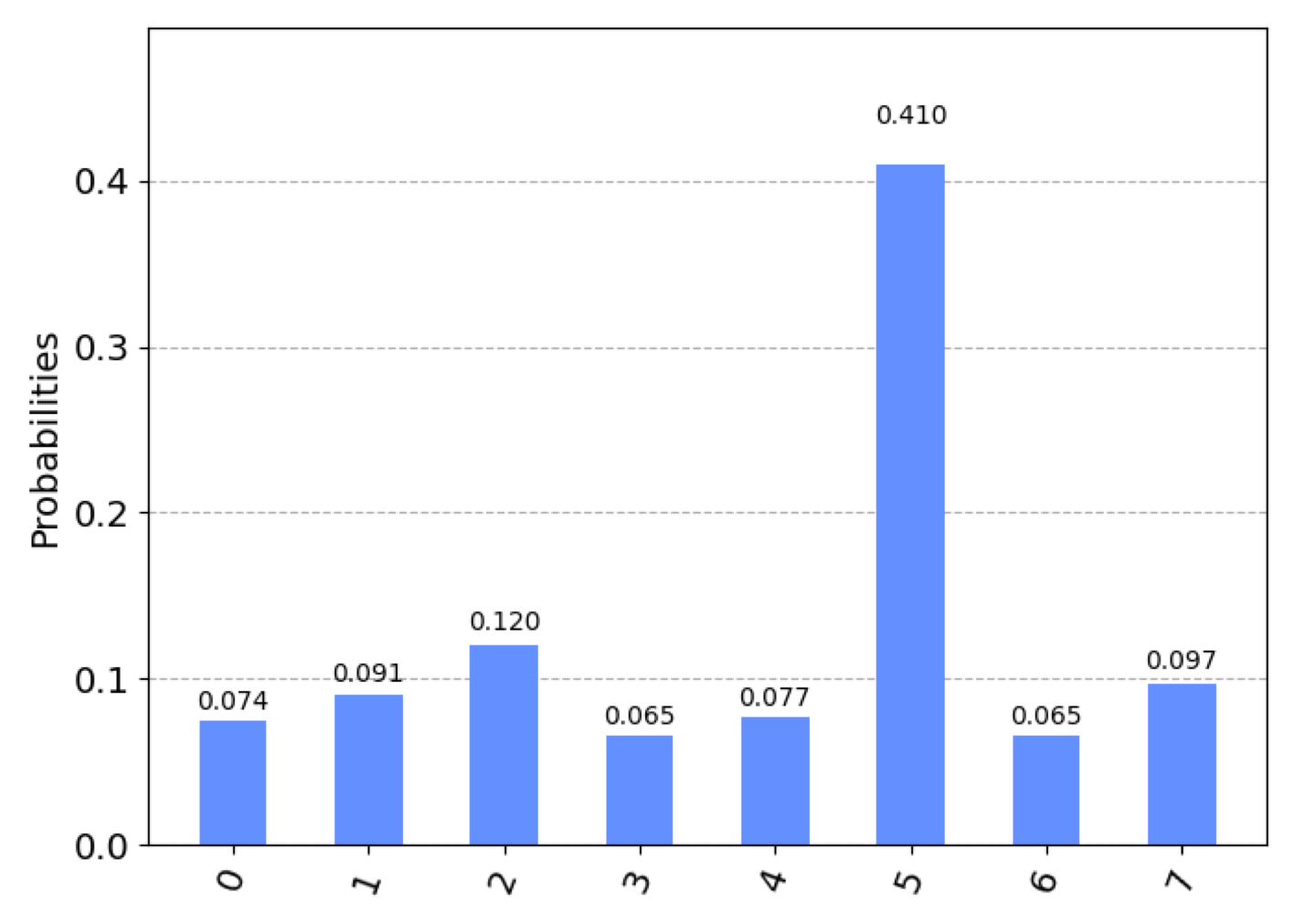}
\caption{\label{fig:grover-3-qubit-set-rx}The result of running a Grover circuit on IBM's \texttt{ibmq\_ourense} backend with 8192 shots, using the \emph{modified} version with $n=3$ qubits, searching for $5$.}
\end{figure} 

\begin{figure}[ht]
\includegraphics[width=7cm]{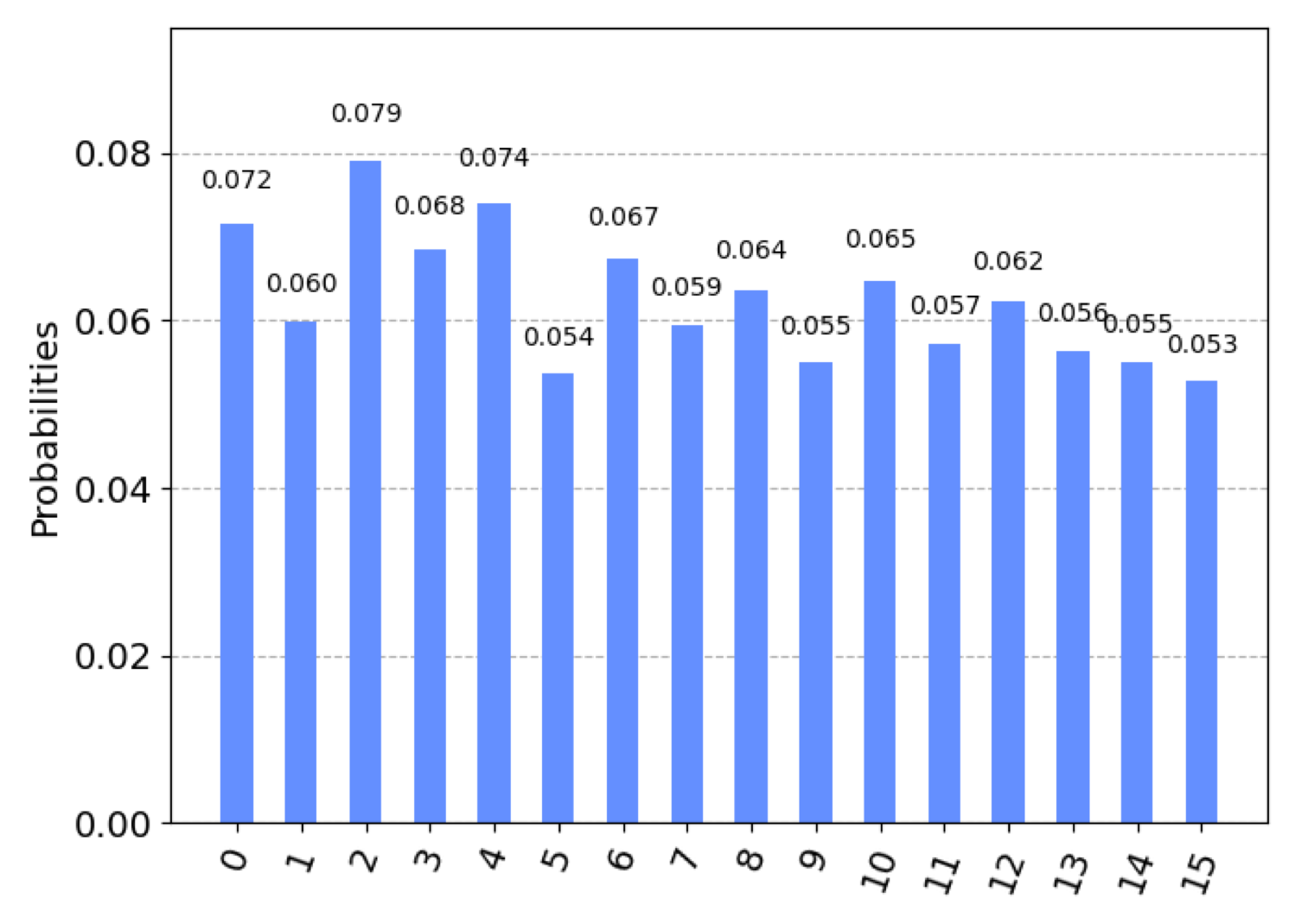}
\caption{\label{fig:grover-4-qubit-set-h}The result of running a Grover circuit on IBM's \texttt{ibmq\_ourense} backend with 8192 shots, using the \emph{standard} version with $n=4$ qubits, searching for $5$.}
\end{figure}

\begin{figure}[ht]
\includegraphics[width=7cm]{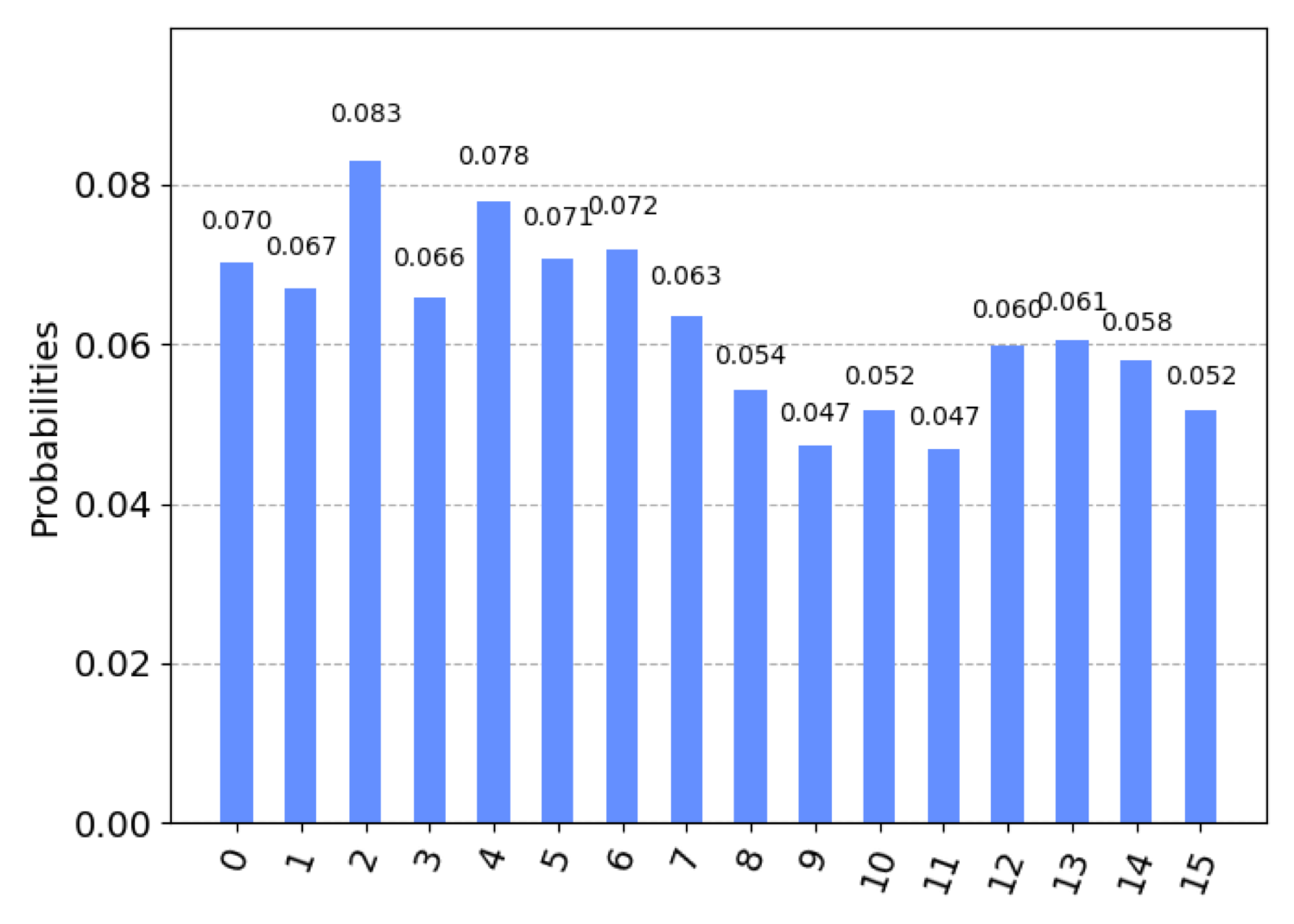}
\caption{\label{fig:grover-4-qubit-set-rx}The result of running a Grover circuit on IBM's \texttt{ibmq\_ourense} backend with 8192 shots, using the \emph{modified} version with $n=4$ qubits, searching for $5$.}
\end{figure}

When using $n=3$ qubits with a marked value of $5$ (shown in Figure~\ref{fig:grover-3-qubit-set-h} and~\ref{fig:grover-3-qubit-set-rx}), even for a high fidelity computer, we see a difference in the measured results.
For $n>3$ qubits, the circuit depth exceeds the limits regarding the coherence time of the quantum hardware currently used. 
For example, the results for $n=4$ qubits are shown in Figures~\ref{fig:grover-4-qubit-set-h} and~\ref{fig:grover-4-qubit-set-rx}.
We receive similar results for the array search example discussed in Section~\ref{sec:grover-list-results}.

\section{\label{sec:related} Related Work}
In this section, we cover related work in the area of Grover's Search algorithm modifications,
particularly those that constitute extensions of the general algorithm with the purpose of
optimizing it.

Boyer, \emph{et al.}~\cite{Boyer1998} provide a tight analysis of Grover’s Search algorithm and propose a formula for computing the probability of finding an element of interest after any given number of iterations of the algorithm.  This can in turn lead to predicting the number of iterations needed to find the given element with high probability. Their analysis also include a model of the algorithm in situations in which the element to be found appears more than once.  For such situations, they provided a new algorithm that works also when the number of solutions is not known in advance. Furthermore, they introduced a new technique for approximate quantum counting and for estimating the number of solutions.  Unlike the solution we present in this paper, their work does not provide an optimization of the inversion-by-the-mean step of Grover's Search.

Mosca~\cite{Mosca1998} introduces a novel interpretation of the eigenvectors and eigenvalues of the iterate operator of Grover's Search.  Their new interpretation leads to novel, optimised algorithm formulations for searching, approximate counting, and amplitude amplification. 

Brassard, \emph{et al.} \cite{Brassard2000} are also interested in extending Grover's Search to perform Amplitude Estimation and apply it to approximate counting.  Unlike our approach, which is based on optimizing the inversion-by-the-mean step, their solution is based on combining ideas from Grover’s and Shor’s quantum algorithms.

\section{\label{sec:conclusion} Conclusion}
We have shown variations in the implementation of the building blocks of the Grover's Search algorithm that use fewer gates, may reduce the number of required iterations, and can lead to overall performance improvements. 

For the common case when a circuit prepares a state by starting with an equal superposition, we have shown a general pattern that replaces Hadamard gates with the simpler $R_X(\frac{\pi}{2})$ gate, allowing for the elimination of a number of $X$ gates. We have used set and array search as examples that fall into this category. 

In geometric terms, one of the takeaways of this paper should be that the state prepared before applying the Amplitude Amplification procedure does not need to be reverted, but instead can be evolved into another state that makes the reflection easier.  

\section*{Disclaimer}
This paper was prepared for information purposes by the Future Lab for Applied Research and Engineering (FLARE) Group of JPMorgan Chase \& Co. and its affiliates, and is not a product of the Research Department of JPMorgan Chase \& Co. 
JPMorgan Chase \& Co. makes no explicit or implied representation and warranty, and accepts no liability, for the completeness, accuracy or reliability of information, or the legal, compliance, tax or accounting effects of matters contained herein.
This document is not intended as investment research or investment advice, or a recommendation, offer or solicitation for the purchase or sale of any security, financial instrument, financial product or service, or to be used in any way for evaluating the merits of participating in any transaction.

IBM, IBM Q, Qiskit are trademarks of International Business Machines Corporation, registered in many jurisdictions worldwide.
Other product or service names may be trademarks or service marks of IBM or other companies.

\section*{Appendix}
\appendix

\section{\label{sec:quantum_dictionary} The Encoding Operator for Array Values}

The encoding operator for array values is in essence the operator described in the quantum-dictionary pattern. An array is a particular case of a dictionary where the keys are used as array indices.

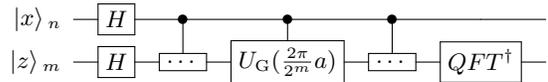
\begin{figure}[ht]
    \mbox{
    \Qcircuit @C=1em @R=0em @!R {
    & \lstick{\ket{x}{}_n} & \gate{H} & \ctrl{1} & \ctrl{1} & \ctrl{1} & \qw & \qw  \\
    & \lstick{\ket{z}{}_m} & \gate{H} & \gate{\text{\ldots}} &  \gate{U_{\text{G}}(\frac{2\pi}{2^{m}}a)} &
    \gate{\text{\ldots}} & \gate{QFT^\dag} & \qw  
    }
}
\caption{\label{fig:operatorA}Circuit for operator $A$.}
\end{figure}

Without including a full description that can be found in our previous work~\cite{Gonciulea2019, Gilliam2019}, if the operator $U_{\text{G}}$ is defined by:

\begin{eqnarray}
U_{\text{G}}(\theta) H^{\otimes m}\ket{0}_m = \frac{1}{\sqrt{2^m}}\sum_{k=0}^{2^m - 1} e^{ik\theta} \ket{k}_m
\end{eqnarray}
then the value encoding operator is of the form $PH$, where $H$ is the Hadamard operator and $P$ is the composition of the sequence of $U_{\text{G}}$ applications followed by ${QFT^\dag}$, as shown in Figure~\ref{fig:operatorA}.

\section{\label{sec:hardware_config} Hardware Configuration}

\begin{figure}[htb]
\includegraphics[width=8cm]{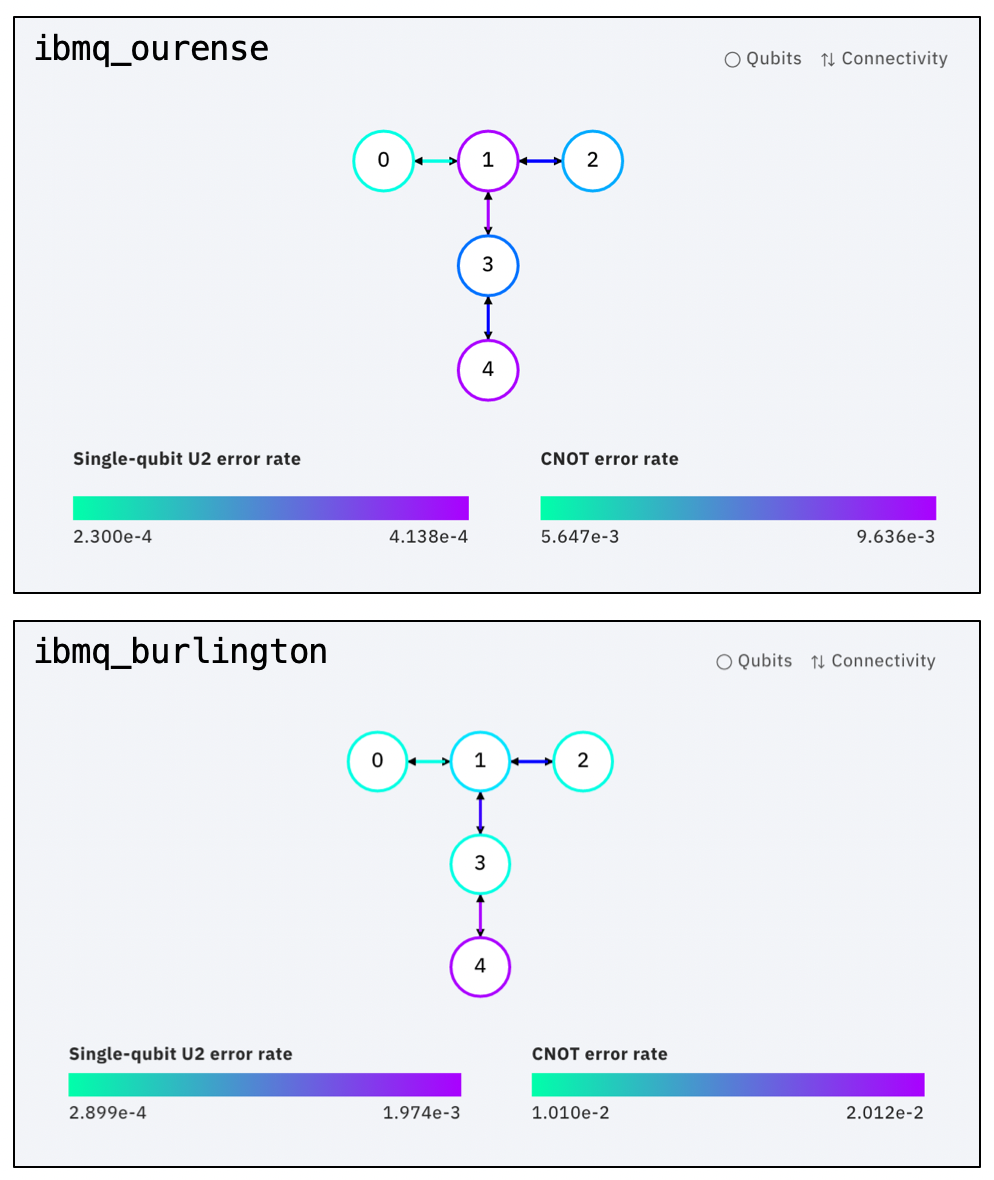}
\caption{\label{fig:hardware-specs}The configuration and error rates for IBM's \texttt{ibmq\_ourense} (top) and \texttt{ibmq\_burlington} (bottom) backend, taken from the IBM Quantum Experience interface at the time of the experiments.}
\end{figure}

In Section~\ref{sec:experimental-results}, we use two quantum devices to run the standard and modified versions of Grover's Search (introduced in Section~\ref{sec:grover-gen}). The configuration and error rates for the devices are listed in Figure~\ref{fig:hardware-specs}.

\section{\label{sec:pixel_based_visualization} Pixel-Based Quantum State Visualization}

\begin{figure}[ht]
\includegraphics[width=3cm]{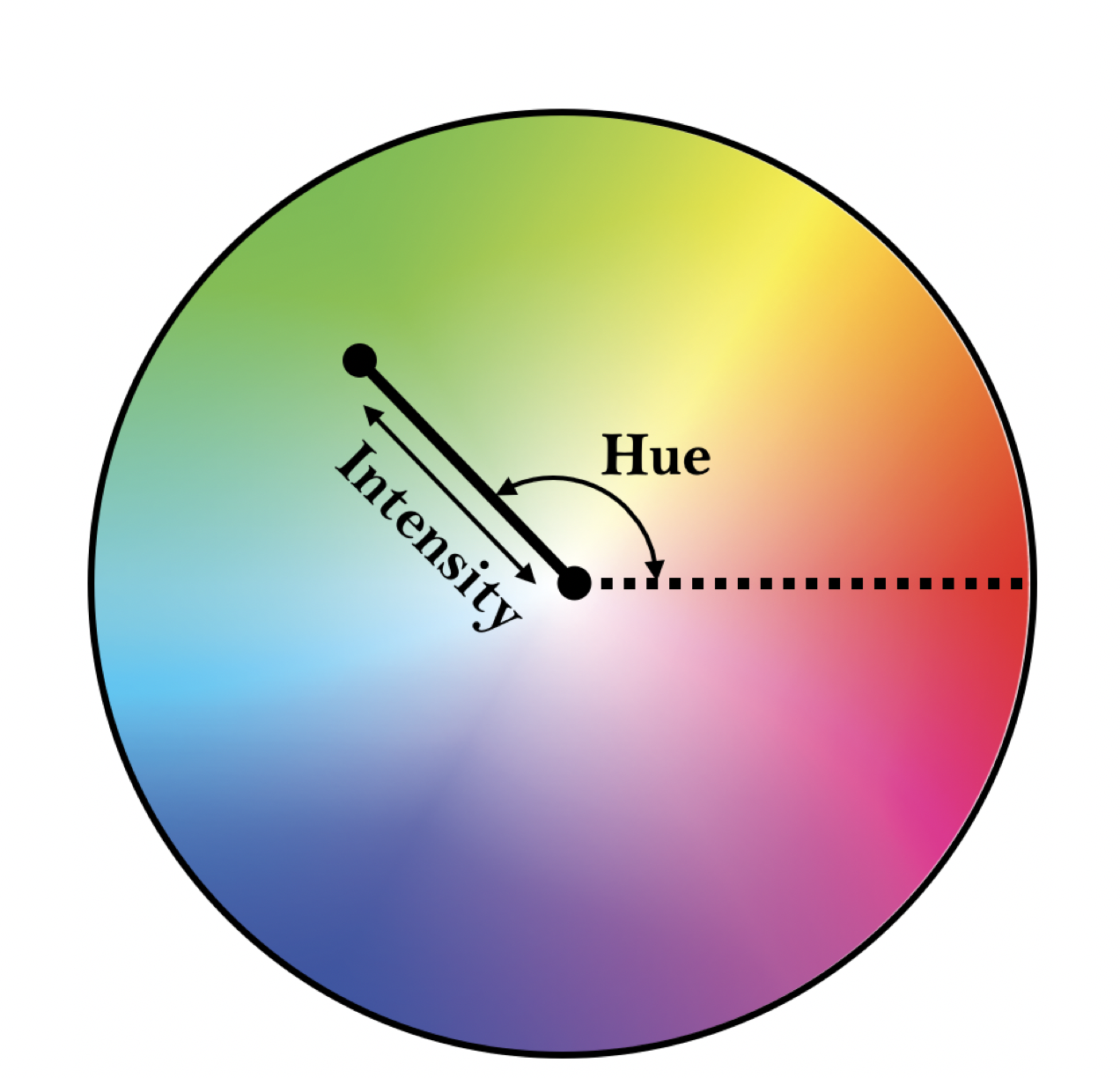}
\caption{\label{fig:color_wheel}A complex number represented in polar form, overlaid onto a color wheel. The phase of the amplitude determines the hue, and the magnitude determines the intensity.} 
\end{figure}

Amplitudes are complex numbers that have a direct correspondence to colors - mapping angles to hues and magnitudes to intensity - as seen in Figure~\ref{fig:color_wheel}.

Using this technique, we can represent the quantum state as a column of pixels, where each pixel corresponds to its respective amplitude. If the computation contains two entangled registers, such as with a quantum dictionary, the visualization is also useful in a tabular form. 

While the mapping of complex numbers to colors is not a new idea (it is commonly used in complex analysis) we find it useful in visualizing quantum state.

\bibliography{main}
 
\end{document}